\documentclass[twocolumn,preprintnumbers,amsmath,amssymb,amsfonts]{revtex4}

\usepackage{graphicx}
\usepackage{mathrsfs}
\usepackage{amssymb}
\usepackage{color}
\usepackage{xcolor}
\usepackage{ulem}
\usepackage{mathtools}
\usepackage{float}
\usepackage{multirow}
\usepackage{diagbox}

\begin{document}
	\title{Phase-sensitive superposition of quantum states}
\author{Xiaotong Wang}
\email{wangxiaotong19@mails.ucas.ac.cn} 

\author{Shunlong Luo}
\email{luosl@amt.ac.cn} 

\author{Yue Zhang}
\email{zhangyue115@amss.ac.cn} 
	\affiliation{State Key Laboratory of Mathematical Sciences, Academy of Mathematics and Systems Science, Chinese Academy of Sciences,  Beijing 100190, China\\
		School of Mathematical Sciences, University of Chinese
		Academy of Sciences, Beijing 100049, China}
	
\begin{abstract}
Although the principle of superposition lies at the heart of quantum mechanics and is the root of almost all quantum phenomena such as coherence and entanglement, its quantification, except for that related to the resource theory of coherence and interference, remains relatively less studied. In this work, we address quantification of superposition from an information-theoretic perspective. We introduce a family of quantifiers of superposition, the phase-sensitive superposition,  by taking into account the phases of amplitudes in the superposition of a fixed basis states (e.g., computational basis states). We establish a conservation relation for the phase-sensitive superposition, which is a kind of complementary relation and is reminiscent of wave-particle duality. We evaluate explicitly the second moment of phase-sensitive superposition and show that it is intrinsically related to  the $l^2$-norm coherence. We characterize the dephasing channel induced by the maximally superposed states. We investigate the minimum and maximum superpositions, reveal their basic properties, and illustrate them through various examples. We further  explore the dynamics of superposition in the Grover search algorithm, and demonstrate  a complementary relation between superposition and success probability of the search algorithm. These results and quantifiers offer  tools for analyzing structural features and implications of quantum superposition. 

		
\vskip 0.3cm
		
	\end{abstract}

	\maketitle
    
\section{Introduction}
The superposition of quantum states in quantum mechanics is not only a mathematical formalism deviating radically from classical realm, but also a physical principle with profound experimental consequences for the microscopic world  \cite{Dira1930}. It is a crucial and nonclassical feature that enables significant  performance advantages in many information
processing tasks \cite{QC1,Niel2000}, such as the Deutsch–Jozsa algorithm \cite{DJ1,DJ2}, the Grover search algorithm \cite{Gro1,Grov19971,Gro2},  the Shor factoring algorithm \cite{Shor1,Shor2}, and various communication protocols \cite{Buhr2010,Guer2016,Tadd2021,Renn2022}, among others. Actually, the superposition principle is the foundation from which nearly all nonclassical quantum resources and phenomena emerge. It serves as the conceptual and mathematical origin for resources such as coherence (superposition in a given basis) \cite{Coh1,Coh2,Coh3,Coh4,Coh5,Coh6,Coh7,Coh8,Coh9,Coh10,Coh11,Coh12,Coh13,Coh14,Coh15,Coh16,Coh17,Coh18,Coh19,Coh20}, entanglement (non-local superposition between subsystems) \cite{Horo2009,Reid2009},  magic (superposition of stabilizer states) \cite{Aaro2004,Brav2005,Dai22,FLS,Feng2025,Fu2025}, bosonic nonclassicality (superposition of coherent states) \cite{luo2019,Zhan2022,NCG1,NCG2,NCG3,NCG4}, and so on. Furthermore, essential quantum characteristics such as wave-particle duality (a manifestation of  superposition of paths) \cite{Woot1979,Engl1996}, and sub-Planck structure (a subtle consequence of superposition) \cite{Zure2001,SubP1,SubP2,SubP3,Luo2024,Tang2025}, are fundamentally linked to quantum superposition. In fact, without the ability of a quantum system to exist in superpositions of multiple states, the vast landscape of quantum technologies would simply not exist. 

Beyond qualitative description of quantum state superposition which has been extensively and intensively discussed,  quantitative characterization has gradually attracted attentions, though still remains relatively few \cite{Aber2006,Theu2017,Chan2017,Toru2021,Teng2023}.   In a seminal work  of {\AA}berg \cite{Aber2006}, two quantifiers of superposition based on analogies with entanglement measures and  unitarily invariant operator norms were proposed. In Ref. \cite{Chan2017}, superposition of a quantum state was quantified by the maximal overlap of the concerned  state with all maximally superposed states.
In Ref. \cite{Theu2017}, a resource-theoretic framework for quantifying superposition in quantum systems was introduced. Recently, superposition was employed to study  texture of quantum states \cite{Pari2024,Zhan2025,Wang2025}, in which a special uniform superposition with vanishing phases plays a basic role.

Superposition is deeply related to coherence and interference, and in the last decade, there  was a vast literature devoted to resource aspects of coherence \cite{Coh1,Coh2,Coh3,Coh4,Coh5,Coh6,Coh7,Coh8,Coh9,Coh10,Coh11,Coh12,Coh13,Coh14,Coh15,Coh16,Coh17,Coh18,Coh19,Coh20}. Apart from these studies and despite the fundamental importance and ubiquity of superposition,  quantitative aspects of superposition itself are still less understood, and it is desirable to quantify superposition from as many angles as possible. Of course, we cannot expect a single quantity to capture all features of superposition, and depending on concrete applications and contexts, different quantifiers may have different usage. In this work, we introduce the notion of phase-sensitive superposition,  study its average and extreme (minimum and maximum) values,  reveal their basic properties, and investigate their implications and applications. 

The remainder of the work is organized as follows. In Sec. II, we introduce phase-sensitive superposition and evaluate the first and the second moments of phase-sensitive superposition explicitly. A conservation law of quantum superposition follows from the average superposition. We characterize the quantum channel induced by the maximally superposed states and relate it to the complete decoherence channel. By evaluating the gradient of phase-sensitive superposition, we cast it  as a quantum resource related to quantum coherence. In  Sec. III, we  investigate the minimum and maximum superpositions. We conclude with a summary in Sec. IV.  To illustrate the quantifiers of superposition, we evaluate the phase-sensitive superposition for some paradigmatic states in Appendix A, further study the dynamics of superposition in Grover search algorithm, and reveal a complementary relation between  the success probability and the maximum superposition in Appendix B.

\section{Quantifying phase-sensitive superposition}
In a $d$-dimensional quantum system described by the Hilbert space $\mathbb{C}^d$ with a fixed orthonormal basis (computational basis) $\{|j\rangle : j\in \mathbb{Z}_d\},$ where $\mathbb{Z}_d =\{0,1,\cdots,d-1\}$ denotes the ring of integers modulo $d$, it is natural to regard the state 
\begin{equation}\label{zero}
|\boldsymbol 0\rangle :=\frac1{\sqrt d}\sum_{j\in\mathbb{Z}_d}|j\rangle
\end{equation}
as a maximally superposed state. Let 
\begin{equation}
F := \frac{1}{\sqrt{d}} \sum_{j, k \in \mathbb{Z}_d} \omega^{j k} |j\rangle \langle k| \label{Fourier}
\end{equation}
be the discrete Fourier transform  on $\mathbb{C}^d$, where $\omega = e^{2\pi i / d}$ is the primitive $d$-th root of unity, then $|\boldsymbol{0}\rangle = F|0\rangle$, which connects the maximally superposed state $|\boldsymbol 0\rangle$ and the computational basis state  $|0\rangle.$ 
 More generally, from both mathematical and  physical points of view, any  superposition state 
\begin{equation}\label{theta}
|\boldsymbol{\theta}\rangle :=\frac1{\sqrt d}\sum_{j\in\mathbb{Z}_d} e^{i\theta_j} |j\rangle
\end{equation}
with arbitrary  phase $\boldsymbol{\theta}=(\theta_0, \theta_1, ...,\theta_{d-1})\in [0,2\pi)^{d}$ can also be equally regarded as  maximally superposed.   
 It can be generated by performing the following  incoherent diagonal unitary transformation 
\begin{equation}
U_{\boldsymbol{\theta}}:= \sum_{j\in \mathbb{Z}_d}e^{i\theta _j}|j\rangle \langle j|=
\begin{pmatrix}
e^{i\theta_0} &0 &\cdots &0\\
0 & e^{i\theta_1} &\cdots &0\\
\cdots &\cdots &\cdots &\cdots\\
0 &0 &\cdots &e^{i\theta_{d-1}}
\end{pmatrix}
\label{DU}
\end{equation}
on $|\boldsymbol 0\rangle$ as
\begin{equation} |\boldsymbol{\theta}\rangle=U_{\boldsymbol{\theta}} |\boldsymbol0\rangle. \label{Ut}
\end{equation}

\vskip 0.2cm

{\bf Definition 1}. For any quantum state (pure or mixed) $\rho$ on $\mathbb{C}^d$, its phase-sensitive superposition is defined as
\begin{equation}\label{def}
S_{\boldsymbol{\theta}}(\rho):=\langle \boldsymbol{\theta}|\rho|\boldsymbol{\theta}\rangle, \qquad \boldsymbol{\theta}\in [0,2\pi)^d,
\end{equation}
which actually is the fidelity between $\rho$ and $|\boldsymbol \theta\rangle\langle\boldsymbol\theta|$ \cite{Joz,Luo2004}. By Eq. (\ref{Ut}), we have
\begin{equation}
    S_\theta(\rho )=\langle \boldsymbol{0}| U_{\boldsymbol\theta}^\dagger  \rho U_{\boldsymbol \theta}|{\boldsymbol 0} \rangle.
    \label{Operation}
\end{equation}

The key point here is to consider the quantity  $S_{\boldsymbol{\theta}}(\rho)$ as a function of the phase $\boldsymbol\theta \in [0, 2\pi)^{d}.$ We will investigate various properties of this function and reveal their physical interpretations and implications.

\vskip 0.2cm

{\bf Proposition 1}. The phase-sensitive superposition $S_{\boldsymbol{\theta}}(\cdot)$ has the following properties.

(1) (Boundedness). $0\leq S_{\boldsymbol{\theta}}(\rho)\leq 1.$ Moreover, $S_{\boldsymbol{\theta}}(\rho)=0$ if and only if   $\rho\in{\rm span}\{ |\boldsymbol{\theta}\rangle\langle \boldsymbol{\theta}|\}^\perp$ (here $A^\perp$ denotes the orthogonal complement of the set $A$ in the Hilbert space $L(\mathbb{C}^d)$ of all linear operators on $\mathbb{C}^d$ endowed with the Hilbert-Schmidt inner product), and 
$S_{\boldsymbol{\theta}}(\rho)= 1$ if and only if $\rho=|\boldsymbol{\theta}\rangle\langle \boldsymbol{\theta}|.$ 


(2) (Positive linearity). $S_{\boldsymbol{\theta}}(\sum_i p_i\rho_i)= \sum_i p_iS_{\boldsymbol{\theta}}(\rho_i)$ for any probability $\{p_i\}$ and quantum states $\rho_i.$

(3) (Monotonicity). $S_{\boldsymbol{\theta}}(\cdot)$ decreases under any quantum channel  $\Lambda$  satisfying $\Lambda(|\boldsymbol\theta\rangle)=|\boldsymbol\theta\rangle$ in the sense that  $S_{\boldsymbol{\theta}}\big(\Lambda(\rho)\big)\leq S_{\boldsymbol{\theta}}(\rho).$ 

(4) (Tensor product). For the product state $\rho_a\otimes\rho_b$ of a bipartite system $\mathbb{C}^{d_a}\otimes \mathbb{C}^{d_b}$ with computational basis $\{|jk\rangle =|j\rangle \otimes |k\rangle: j\in \mathbb{Z}_{d_a}, k\in \mathbb{Z}_{d_b}\},$ we have
$$S_{\boldsymbol{\theta}_{ab}}(\rho_a\otimes\rho_b)=S_{\boldsymbol{\theta}_a}(\rho_a)S_{\boldsymbol{\theta}_b}(\rho_b),$$ 
where   $\boldsymbol{\theta}_x=(\theta _{x0}, \theta _{x1}, \cdots, \theta _{x(d_x-1)})$  ($x=a,b$), and 
$$|\boldsymbol{\theta}_{ab} \rangle =|\boldsymbol{\theta}_a\rangle \otimes  |\boldsymbol{\theta}_b \rangle = \frac1{\sqrt {d_ad_b}}\sum_{j\in\mathbb{Z}_{d_a},k\in\mathbb{Z}_{d_b}} e^{i(\theta_{aj}+\theta_{bk})} |jk\rangle$$ 
with $\boldsymbol{\theta}_{ab}=(\theta_{a0}+\theta_{b0},\theta_{a0}+\theta_{b1},\dots, \theta_{a(d_a-1)}+\theta_{b(d_b-1)}).$

\vskip 0.2cm

 All the above properties follow rather directly from the definition in Eq. (\ref{def}).

We remark that $S_{\boldsymbol0}(\rho) ={\rm tr} ( \rho |\boldsymbol0\rangle\langle \boldsymbol0|)$ is essentially related to the so-called  quantum state texture, $-{\rm ln}S_{\boldsymbol 0}(\rho), $ as introduced in Ref. \cite{Pari2024}.

Now, we examine the average properties of the phase-sensitive superposition $S_{\boldsymbol\theta}(\cdot )$ over the whole parameter space $\boldsymbol\theta\in [0,2\pi)^d$. 

\vskip 0.2cm

{\bf Proposition 2}.  In a $d$-dimensional quantum system $\mathbb{C}^d$, the following results hold
\begin{align}
 \int_{[0,2\pi)^d} |\boldsymbol{\theta}\rangle\langle \boldsymbol{\theta}| \frac{{\rm d}\boldsymbol{\theta}}{(2\pi)^d} & = \frac{{\mathbf 1}_d}{d}, \label{1M}\\
 \int_{[0,2\pi)^d}|\boldsymbol{\theta}\rangle\langle \boldsymbol{\theta}|^{\otimes 2}
\frac{{\rm d}\boldsymbol{\theta}}{(2\pi)^d} &= \frac{1}{d^2} \Big ( \mathbf 1_{d^2} + S - \sum_j |j\rangle\langle j| \otimes |j\rangle\langle j| \Big ),\label{2M}
\end{align}
where $\mathbf 1_{d}$ is the identity operator on $\mathbb{C}^d,$   $\mathbf 1_{d^2}$ is the identity operator on $\mathbb{C}^d \otimes \mathbb{C}^d,$ 
and $S$ is the swap operator on $\mathbb{C}^d \otimes \mathbb{C}^d$ determined  by $S(|j\rangle \otimes |k\rangle) = |k\rangle \otimes |j\rangle,\ \forall j,k\in\mathbb Z_d$.

\vskip 0.2cm
Eq. (\ref{1M}) already implicitly appeared  in Ref. \cite{Chan2017}. Here we provide a brief proof for reader's convenience. Noting Eq. (\ref{theta}), we have
$$
|\boldsymbol{\theta}\rangle\langle \boldsymbol{\theta}|
= \frac{1}{d}\sum_{j,k\in\mathbb Z_d} e^{i(\theta_j-\theta_k)} |j\rangle\langle k|,
$$
and then
$$
\begin{aligned}
&\int_{[0,2\pi)^d} |\boldsymbol{\theta}\rangle\langle \boldsymbol{\theta}|
\frac{{\rm d}\boldsymbol{\theta}}{(2\pi)^d}\\
&= \frac{1}{d}\sum_{j,k\in\mathbb Z_d} |j\rangle\langle k|
\int_{[0,2\pi)^d} e^{i(\theta_j-\theta_k)}
\frac{{\rm d}\boldsymbol{\theta}}{(2\pi)^d} \\
&= \frac{1}{d}\sum_{j,k\in\mathbb Z_d} |j\rangle\langle k| \delta_{jk}
= \frac{{\mathbf 1}_d}{d}.
\end{aligned}
$$
To establish Eq. (\ref{2M}), noting that
$$
|\boldsymbol{\theta}\rangle\langle \boldsymbol{\theta}|^{\otimes 2} 
= \frac{1}{d^2} \sum_{j,k,l,m\in\mathbb Z_d} e^{i(\theta_j-\theta_k + \theta_l-\theta_m)} |j\rangle\langle k| \otimes |l\rangle\langle m|,
$$
we have
$$
\begin{aligned}
&\int_{[0,2\pi)^d} |\boldsymbol{\theta}\rangle\langle \boldsymbol{\theta}|^{\otimes 2}
\frac{{\rm d}\boldsymbol{\theta}}{(2\pi)^d} \\
&= \frac{1}{d^2} \sum_{j,k,l,m\in\mathbb Z_d} |j\rangle\langle k| \otimes |l\rangle\langle m|
\int_{[0,2\pi)^d} e^{i(\theta_j-\theta_k + \theta_l-\theta_m)}
\frac{{\rm d}\boldsymbol{\theta}}{(2\pi)^d} \\
&= \frac{1}{d^2} \sum_{j,k,l,m\in\mathbb Z_d} |j\rangle\langle k| \otimes |l\rangle\langle m|
\bigl( \delta_{jk}\delta_{lm} + \delta_{jm}\delta_{kl} - \delta_{jk}\delta_{lm}\delta_{jl} \bigr) \\
&= \frac{1}{d^2} \Bigl( \sum_{j,l\in\mathbb Z_d} |j\rangle\langle j| \otimes |l\rangle\langle l| + \sum_{j,l\in\mathbb Z_d} |j\rangle\langle l| \otimes |l\rangle\langle j| \\
&\qquad \quad - \sum_{j\in\mathbb Z_d} |j\rangle\langle j| \otimes |j\rangle\langle j| \Bigr) \\
&= \frac{1}{d^2} \Bigl( \mathbf 1_{d^2} + S - \sum_{j\in\mathbb Z_d} |j\rangle\langle j| \otimes |j\rangle\langle j| \Bigr).
\end{aligned}
$$

\vskip 0.2cm

In view of Eq. (\ref{1M}), we conclude that
$$
\bigl\{d|\boldsymbol{\theta}\rangle\langle \boldsymbol{\theta}| \;:\; \boldsymbol{\theta}\in [0,2\pi)^d \bigr\}
$$ 
is a positive operator valued measure (POVM), which is equivalent to saying that 
\begin{equation}
\boldsymbol\Theta=\{|\boldsymbol\theta\rangle : \boldsymbol\theta\in[0,2\pi)^d\} \label{The}
\end{equation} 
is a quantum 1-design \cite{Zauner2011}.

From Eq. (\ref{2M}), we see however that $\boldsymbol{\Theta}$ is not a quantum 2-design. Recall the standard expression for the second moment of Haar-uniform pure states over $\mathbb{C}^d$, which defines a quantum 2-design  as \cite{Zauner2011}
\begin{equation}
\int_{|\psi\rangle\in\mathbb{C}^d} |\psi\rangle\langle \psi|^{\otimes 2} \, d\psi
= \frac{1}{d(d+1)} \bigl( \mathbf{1}_{d^2} + S \bigr). 
\label{quantum-2-design}
\end{equation}
In contrast, the integral \eqref{2M} over $\boldsymbol{\Theta}$ contains an extra term, $
-\sum_{j\in\mathbb{Z}_d} |j\rangle\langle j| \otimes |j\rangle\langle j|.
$
This operator projects onto the diagonal product subspace
$
\operatorname{span}\bigl\{ |j\rangle\otimes|j\rangle : j\in \mathbb{Z}_d \bigr\},
$
which is dependent on the chosen basis $\{|j\rangle: j\in \mathbb{Z}_d\}$.
The appearance of this basis-dependent term reflects the fact that superposition is defined relative to a fixed reference basis. Consequently, the averaging over $\boldsymbol{\Theta}$ cannot achieve the fully basis-invariant property required of a genuine 2-design.

From Proposition 2, we can derive  the first and second moments of $S_{\boldsymbol{\theta}}(\rho)$ as follows.

\vskip 0.2cm
{\bf Proposition 3}.
In a $d$-dimensional quantum system $\mathbb{C}_d,$ the phase-sensitive superposition $S_{\boldsymbol{\theta}}(\rho)$ satisfies
\begin{align}
    \int_{[0, 2\pi)^{d}} S_{\boldsymbol{\theta}}(\rho)\frac{{\rm d}\boldsymbol{\theta} }{(2\pi)^d} &=\frac 1d,     \label{conservation law}\\
    \int_{[0, 2\pi)^{d}} S_{\boldsymbol{\theta}}^2 (\rho)\frac{{\rm d}\boldsymbol{\theta} }{(2\pi)^d}&=\frac{1}{d^2}\bigr(1+C_{l^2}(\rho)\bigr),\label{SM}
    \end{align}
    where 
    \begin{equation}
    C_{l^2}(\rho)=\sum_{ j \neq k} |\langle j|\rho |k\rangle |^2 = \sum_{ j \neq k} |\rho_{jk}|^2 \label{Coherence}
    \end{equation}
     is the $l^2$-norm coherence of $\rho$ relative to the fixed basis $\{|j\rangle: j\in \mathbb{Z}_d\}$.
    

\vskip 0.2cm

Eq. (\ref{conservation law}) follows readily from (\ref{1M}), and  Eq. (\ref{SM}) follows from  
$$
\begin{aligned}
& \int_{[0, 2\pi)^{d}} S_{\boldsymbol{\theta}}^2 (\rho)\frac{{\rm d}\boldsymbol{\theta} }{(2\pi)^d}\\
&=\mathrm{tr}\!\Big(\rho^{\otimes 2} \int_{[0,2\pi)^d}|\boldsymbol{\theta}\rangle\langle \boldsymbol{\theta}|^{\otimes 2}
\frac{{\rm d}\boldsymbol{\theta}}{(2\pi)^d}\Big )\\
&= \frac{1}{d^2} \mathrm{tr} \Bigl( \rho^{\otimes 2} \bigr( \mathbf 1_{d^2} + S - \sum_{j\in\mathbb Z_d} |j\rangle\langle j| \otimes |j\rangle\langle j| \bigr) \Bigr)\\
&= \frac{1}{d^2} \Bigl( (\mathrm{tr} \rho)^2 + \mathrm{tr}(\rho^2) - \sum_{j\in\mathbb Z_d} \langle j|\rho |j\rangle^2 \Bigr)\\
&=\frac{1}{d^2} \Big (1+ \sum_{j \neq k} |\rho_{jk}|^2 \Big ).
\end{aligned}
$$
The third equality invokes the property  
$$\mathrm{tr}\big ((A\otimes B) S\big )=\mathrm{tr}(AB)$$ of the swap operator $S$  for any operators  $A$ and $B$ on $\mathbb{C}^d$.

Eq. \eqref{conservation law} reveals a conservation relation 
for phase-sensitive superposition over the entire parameter space: 
If a state yields a large value of $ S_{\boldsymbol{\theta}}(\rho) $ for 
some phase $ \boldsymbol{\theta} $, it must be compensated by smaller 
values for other phases.

Eq. \eqref{SM} establishes a fundamental connection between phase-sensitive superposition and quantum coherence. It provides an operational interpretation of the $l^2$-norm coherence $C_{l^2}(\rho)$ as the average squared fidelity between the state $\rho$ and the ensemble of all maximally superposed states $|\boldsymbol{\theta}\rangle$ (with respect to the chosen computational basis). This relation yields a practical protocol for estimating $C_{l^2}(\rho)$ without full quantum state tomography. 
In view of $|\boldsymbol{0}\rangle = F|0\rangle$ with $F$ the discrete Fourier transform defined by Eq. (\ref{Fourier}), Eq.~\eqref{Operation} can be rewritten as
$$
S_{\boldsymbol{\theta}}(\rho) = \langle 0| F^\dagger U_{\boldsymbol{\theta}}^\dagger \rho \, U_{\boldsymbol{\theta}} F |0\rangle.
$$
The experimental protocol is thus operationally well-defined: Apply a random diagonal unitary $U_{\boldsymbol{\theta}}$, followed by the discrete Fourier transform $F$, and perform a computational-basis measurement on $|0\rangle$. By repeating the procedure for many independent random phase vectors $\{\boldsymbol{\theta}_k\}$ and averaging the resulting squared probabilities $S_{\boldsymbol{\theta}_k}^2(\rho)$, one obtains a direct statistical estimate of $C_{l^2}(\rho)$ via Eq.~\eqref{SM}.

\vskip 0.1cm

Another perspective on phase-sensitive superposition is provided by investigating the quantum channel
\begin{equation}
\mathcal{E}(\rho):=d\int_{[0, 2\pi)^{d}}|\boldsymbol{\theta}\rangle\langle \boldsymbol{\theta}|\rho |\boldsymbol{\theta}\rangle\langle \boldsymbol{\theta}| \frac{{\rm d}\boldsymbol{\theta} }{(2\pi)^d} \label{Cht}
\end{equation}
induced by $\boldsymbol{\Theta}.$  Recall that  the complete decoherence channel on a $d$-dimensional system is defined as 
\begin{equation}
{\cal D}(\rho ):=\sum_{j} \Pi _j\rho \Pi_j=\sum_{j} \rho_{jj}  |j\rangle\langle j|, \label{cdc}
\end{equation}
where $\Pi _j=|j\rangle \langle j|$ and $\rho =(\rho _{ij}) $ is in the matrix form relative to the fixed basis $\{|j\rangle :j\in \mathbb{Z}_d\}.$

\vskip 0.2cm

{\bf Proposition 4}.  The channel $\mathcal{E}$ defined by Eq. (\ref{Cht}) is related to the complete decoherence channel ${\cal D}$ defined by Eq. (\ref{cdc}) as
\begin{equation}
    \mathcal{E}(\rho)= \frac{1}{d}\big (\mathbf{1}+\rho-{\cal D}(\rho )\big ).      \label{channel}
\end{equation}

\vskip 0.2cm

The above result follows from 
$$
\langle \boldsymbol{\theta} | \rho | \boldsymbol{\theta} \rangle = \frac{1}{d} \sum_{j',k'\in\mathbb{Z}_d} \rho_{k'j'} e^{i(\theta_{j'} - \theta_{k'})},
$$
we have
\begin{align*}
&d\int_{[0, 2\pi)^{d}}|\boldsymbol{\theta}\rangle\langle \boldsymbol{\theta}|\rho |\boldsymbol{\theta}\rangle\langle \boldsymbol{\theta}| \frac{{\rm d}\boldsymbol{\theta} }{(2\pi)^d}\\
   &= d \int  \Big ( \frac{1}{d} \sum_{j,k\in\mathbb Z_d} e^{i(\theta_j - \theta_k)} |j\rangle\langle k| \Big )
   \times \\
   & \qquad \quad  \Big ( \frac{1}{d} \sum_{j',k'\in\mathbb{Z}_d} \rho_{k'j'} e^{i(\theta_{j'} - \theta_{k'})} \Big )\frac{d\boldsymbol{\theta}}{(2\pi)^d}\\
   &=\frac{1}{d} \sum_{j,k,j',k'\in\mathbb Z_d} \rho_{k'j'} |j\rangle\langle k| \int  e^{i(\theta_j - \theta_k + \theta_{j'} - \theta_{k'})}\frac{d\boldsymbol{\theta}}{(2\pi)^d}\\
   &=\frac{1}{d} \sum_{j,k,j',k'\in\mathbb Z_d} \rho_{k'j'} |j\rangle\langle k| (\delta_{jk}\delta_{j'k'} + \delta_{jk'}\delta_{kj'} - \delta_{jk}\delta_{j'k'}\delta_{jj'})\\
   &=\frac{1}{d} \Big ( \sum_{j,m\in\mathbb Z_d} \rho_{mm}  |j\rangle\langle j| + \sum_{j,k\in\mathbb Z_d} \rho_{jk}  |j\rangle\langle k| - \sum_{j\in\mathbb Z_d} \rho_{jj}  |j\rangle\langle j| \Big )\\
   &= \frac{1}{d}\Big (\mathbf{1}+\rho-\sum_{j\in\mathbb Z_d} \rho_{jj}  |j\rangle\langle j| \Big ),
\end{align*}
which is the desired result.

Thus, the channel $\mathcal{E}$ preserves the coherence information encoded in off-diagonal elements, while mapping the diagonal elements to a uniform distribution. This shows that $\mathcal{E}$ is complementary to the completely decoherence channel $\mathcal{D}$, which eliminates off-diagonal elements and preserves the diagonal ones.

We now examine the gradient of $ S_{\boldsymbol{\theta}}(\rho) $, which measures how sensitive the state $ \rho $ is to local phase changes. Recall that
$$
S_{\boldsymbol{\theta}}(\rho) = \langle \boldsymbol{\theta} | \rho | \boldsymbol{\theta} \rangle = \frac{1}{d} \sum_{j,k} \rho_{jk} e^{-i(\theta_j - \theta_k)},
$$
where $ \rho_{jk} = \langle j | \rho | k \rangle = |\rho_{jk}|e^{i\phi_{jk}}.$ For any fixed state $ \rho $, regarding $S_{\boldsymbol{\theta}}(\rho) $ as a function of the phase $ \boldsymbol{\theta}$ and taking the partial derivative with respect to the $ j $-th component $ \theta_j $, we obtain
\begin{align*}
  \frac{\partial S_{\boldsymbol\theta}(\rho)}{\partial \theta_j} &= \frac{i}{d} \sum_{k} \big ( \rho_{kj} e^{i(\theta_j - \theta_k)} - \rho_{jk} e^{-i(\theta_j - \theta_k)} \big ) \\
  &= \frac{2}{d} \sum_{k \neq j} |\rho_{jk}| \sin\!\big ( \phi_{jk} - (\theta_j - \theta_k) \big ).
\end{align*}
The gradient vector of $ S_{\boldsymbol\theta} (\rho) $ with respect to $ \boldsymbol{\theta} $ is defined as
$$
\nabla S_{\boldsymbol{\theta}}(\rho)=\left( \frac{\partial S_{\boldsymbol\theta}(\rho)}{\partial \theta_0}, \cdots,  \frac{\partial S_{\boldsymbol\theta}(\rho)}{\partial \theta_{d-1}} \right).
$$
When $ \nabla S_{\boldsymbol{\theta}}(\rho) = 0 $, $ \boldsymbol{\theta} $ corresponds to either a local maximum, a local minimum, or a saddle point of $ S_{\boldsymbol{\theta}}(\rho) $. However, solving this system of nonlinear equations is generally not straightforward. The following proposition presents some fundamental properties of the gradient and the Hessian of $S_{\boldsymbol{\theta}}(\rho)$, both in the pointwise and average senses.

\vskip 0.2cm

\textbf{Proposition 5.} For any quantum state $\rho$ in  a $d$-dimensional system $\mathbb{C}^d$, we have
\begin{align}
  &\sum_{j \in \mathbb{Z}_d} \frac{\partial S_{\boldsymbol\theta}(\rho)}{\partial \theta_j} = 0, \ \ \forall \ \boldsymbol{\theta} = (\theta_0, \dots, \theta _{d-1}) \in [0, 2\pi)^d,  \label{T1} \\
  &\int_{[0,2\pi)^d} \sum_{j \in \mathbb{Z}_d} \left( \frac{\partial S_{\boldsymbol\theta}(\rho)}{\partial \theta_j} \right)^2 \frac{\mathrm{d}\boldsymbol{\theta}}{(2\pi)^d} = \frac{2}{d^2} C_{l^2}(\rho), \label{T2} \\
  &\int_{[0,2\pi)^d} \sum_{j,k \in \mathbb{Z}_d} \left( \frac{\partial^2 S_{\boldsymbol\theta}(\rho)}{\partial \theta_j \partial \theta_k} \right)^2 \frac{\mathrm{d}\boldsymbol{\theta}}{(2\pi)^d} = \frac{4}{d^2} C_{l^2}(\rho), \label{T3}
\end{align}
where $ C_{l^2}(\rho)$ is the $ l^2$-norm coherence defined by Eq. (\ref{Coherence}).

\vskip 0.2cm

Eq. \eqref{T1} follows from
$$
\sum_{j \in \mathbb{Z}_d} \frac{\partial S_{\boldsymbol\theta}(\rho)}{\partial \theta_j} = \frac{2}{d} \sum_{ j \neq k} |\rho_{jk}| \sin(\phi_{jk} - \theta_j + \theta_k) = 0.
$$
The last equation holds because $\rho_{jk}=\bar \rho_{kj}$ and $\phi_{jk}=-\phi_{kj}$, causing the terms for each pair $(j,k)$ and $(k,j)$ to cancel.  Eq. \eqref{T1} reflects the global-phase invariance of the quantum state. Specifically, if we shift all phases uniformly by the same amount $ \delta $, i.e., $ (\theta_0, \dots, \theta_{d-1}) \mapsto (\theta_0 + \delta, \dots, \theta_{d-1} + \delta) $, the state $ |\boldsymbol{\theta}\rangle $ remains unchanged up to a global phase, and therefore $ S_{\boldsymbol{\theta}}(\rho) $ is invariant under such a shift. 

Eq. \eqref{T2} follows from
\begin{align*}
  &\int_{[0,2\pi)^d} \sum_{j \in \mathbb{Z}_d} \left( \frac{\partial S_{\boldsymbol\theta}(\rho)}{\partial \theta_j} \right)^2 \frac{\mathrm{d}\boldsymbol{\theta}}{(2\pi)^d} \\
  &= \frac{4}{d^2} \sum_{ j \neq k} |\rho_{jk}|^2 \int_{[0,2\pi)^d} \sin^2\!\left( \phi_{jk} - \theta_j + \theta_k \right) \frac{\mathrm{d}\boldsymbol{\theta}}{(2\pi)^d} \\
  &= \frac{2}{d^2} \sum_{ j \neq k} |\rho_{jk}|^2.
\end{align*}

To derive Eq. \eqref{T3}, we compute
\begin{align*}
  &\frac{\partial^2 S_{\boldsymbol\theta}(\rho)}{\partial \theta_j \partial \theta_k} = \frac{2}{d} |\rho_{jk}| \cos\big( \phi_{jk} - (\theta_j - \theta_k) \big), \quad j \neq k. \\
  &\frac{\partial^2 S_{\boldsymbol\theta}(\rho)}{\partial \theta_j ^2} = -\frac{2}{d} \sum_{ k \neq j} |\rho_{jk}| \cos\big( \phi_{jk} - (\theta_j - \theta_k) \big).
\end{align*}
Therefore, for $ j \neq k $,
\begin{align*}
  &\int_{[0,2\pi)^d} \left( \frac{\partial^2 S_{\boldsymbol\theta}(\rho)}{\partial \theta_j \partial \theta_k} \right)^2 \frac{\mathrm{d}\boldsymbol{\theta}}{(2\pi)^d} \\
  &= \frac{4}{d^2} |\rho_{jk}|^2 \int_{[0,2\pi)^d} \cos^2\big( \phi_{jk} - (\theta_j - \theta_k) \big) \frac{\mathrm{d}\boldsymbol{\theta}}{(2\pi)^d} \\
  &= \frac{2}{d^2} |\rho_{jk}|^2.
\end{align*}
Similarly,
$$
\int_{[0,2\pi)^d} \left( \frac{\partial^2 S_{\boldsymbol\theta}(\rho)}{\partial \theta_j ^2} \right)^2 \frac{\mathrm{d}\boldsymbol{\theta}}{(2\pi)^d} = \frac{2}{d^2} \sum_{k \in \mathbb{Z}_d, k \neq j} |\rho_{jk}|^2.
$$
Adding the above two equations together yields Eq. \eqref{T3}.

Eqs. \eqref{T2} and \eqref{T3} demonstrate that the average norm of both the gradient and the Hessian of $S_{\boldsymbol{\theta}}(\rho)$ are essentially the coherence of the state $\rho$. This establishes a direct relationship between phase sensitivity and quantum coherence.

\section{Extreme values of phase-sensitive superposition}

We now study the extreme values of phase-sensitive superposition $S_{\boldsymbol{\theta}}(\rho )$ when the phase $\boldsymbol{\theta}$ varies.

\subsection{Minimal superposition}

The minimal value of phase-sensitive superposition is defined as  
\begin{equation}
S_{\min}(\rho):=\min_{\boldsymbol{\theta}} \langle \boldsymbol{\theta}|\rho|\boldsymbol{\theta}\rangle,
\end{equation}
which can also be expressed as
$$S_{\min}(\rho)=\min_{\boldsymbol{\theta}}  \langle \boldsymbol{0}| U_{\boldsymbol \theta} \rho U_{\boldsymbol \theta}^\dag|{\boldsymbol 0} \rangle$$
with $U_{\boldsymbol\theta}$ defined by Eq. (\ref{DU}).

\vskip 0.2cm

{\bf Proposition 6}. The minimal superposition $S_{\min}(\cdot)$ has the following properties.

(1) (Boundedness). $0\leq S_{\rm min}(\rho)\leq 1/d.$ The upper bound 
$S_{\rm min}(\rho)=1/d$ is reached if and only if $\rho$ is diagonal in the base $\{|j\rangle:j\in\mathbb Z_d\}.$


(2) (Concavity). $S_{\rm min}(\sum_i p_i\rho_i)\geq \sum_i p_iS_{\rm min}(\rho_i)$ for any probability $\{p_i\}$ and quantum states $\rho_i.$ 


(3) (Tensor product bound). For any quantum states $\rho_a$ and $\rho_b$ of two systems $a$ and $b$,  
$$S_{\rm min}(\rho_a \otimes \rho_b) \leq S_{\rm min} (\rho_a)S_{\rm min}(\rho_b).$$

(4) (Invariance). The minimal superposition is invariant under diagonal unitary transformation in the sense that
$$S_{\rm min}(U\rho U^\dag)=S_{\rm min}(\rho )$$
for any unitary  $U$ in the form of Eq. (\ref{DU}).

\vskip 0.2cm

The non-negativity in item (1) is apparent from the definition of $S_{\min}(\rho),$ while the upper bound follows from the argument that if $S_{\min}(\rho)>1/d,$ then 
$$\int _{[0,2\pi )^d} S_{\boldsymbol\theta} (\rho) \frac{\mathrm{d}\boldsymbol{\theta}}{(2\pi)^d} \geq  \int _{[0,2\pi )^d} S_{\min}(\rho)  \frac{\mathrm{d}\boldsymbol{\theta}}{(2\pi)^d} >\frac 1d,$$
which  contradicts to Eq. (\ref{conservation law}) in Proposition 3. Item (2) follows from  
$$\min_{\boldsymbol\theta} \sum_i p_i\langle \boldsymbol{\theta}|\rho_i|\boldsymbol{\theta}\rangle\geq \sum_i p_i \min_{\boldsymbol\theta} \langle \boldsymbol{\theta}|\rho_i|\boldsymbol{\theta}\rangle. $$
Item (3) follows from
\begin{align*}
\min_{\boldsymbol{\theta}} \langle \boldsymbol{\theta}|\rho_a \otimes \rho_b|\boldsymbol{\theta}\rangle &\leq \min_{{\boldsymbol{\theta}}_a,{\boldsymbol{\theta}}_b} \langle \boldsymbol{\theta}_a \otimes \boldsymbol{\theta}_b|\rho_a \otimes \rho_b|\boldsymbol{\theta}_a \otimes \boldsymbol{\theta}_b\rangle \\
&= \min_{\boldsymbol{\theta}_a} \langle \boldsymbol{\theta}_a |\rho_a |\boldsymbol{\theta}_a \rangle \min_{\boldsymbol{\theta}_b} \langle \boldsymbol{\theta}_b |\rho_b |\boldsymbol{\theta}_b \rangle.
\end{align*} 
Item (4) follows from direct verification.

\vskip 0.15cm

The minimal superposition $S_{\min}(\cdot)$ quantifies the irreducible superposition of $\rho$. It captures the worst-case alignment between $\rho$ and any state in $\boldsymbol\Theta$: A residual presence that cannot be removed by any choice of $|\boldsymbol{\theta}\rangle$.

To gain insight into the meaning of $S_{\min}(\cdot)$, we consider the case where $\rho =|\psi \rangle \langle \psi |$ is a pure state on $\mathbb{C}^d$ with
\begin{equation} \label{pure}
|\psi \rangle = \sum_{j\in \mathbb Z_d} a_j |j\rangle,\qquad \sum_{j\in \mathbb Z_d} |a_j |^2=1,
\end{equation}
then $$S_{\min}(|\psi\rangle)=\min _{\boldsymbol\theta} \Big|\frac{1}{\sqrt{d}} \sum_{j\in \mathbb Z_d} e^{-i\theta_j}a_j\Big|^2.$$
We have the following result.

\vskip 0.2cm

{\bf Proposition 7}. 
For any $|\psi \rangle= \sum_{j\in \mathbb Z_d} a_j |j\rangle \in \mathbb{C}^d$, 
\begin{equation}\label{Smin}
   S_{\min}(|\psi\rangle) =\frac{1}{d} \Big(\max\Big\{0,  2 \max_j |a_j| - \sum_{j\in \mathbb Z_d} |a_j| \Big\}\Big)^2.
\end{equation}
Let $a_j = |a_j| e^{i\gamma _j}$ be the polar decomposition. The minimum is attained as follows.

(1) If $2\max_j |a_j| > \sum_{j} |a_j|$, then the minimum is nonzero and is achieved by the phase vector
    $$
    \boldsymbol\theta = (\gamma _0, \dots, \gamma _{j_0-1},\; \gamma _{j_0}+\pi,\; \gamma _{j_0+1}, \dots, \gamma _{d-1}),
    $$
    where $j_0$ is any index satisfying $|a_{j_0}| = \max_j |a_j|$.
    
(2) If $2\max_j |a_j| \le \sum_{j} |a_j|$, then $S_{\min}(|\psi\rangle)=0$, and the zero value is attained for any $\boldsymbol\theta$ satisfying
    $$
     \sum_{j \neq j_0} e^{-i\theta_j} a_j  = |a_{j_0}|, \qquad  \theta_{j_0}=\gamma _{j_0}+\pi, 
    $$
    where again $|a_{j_0}| = \max_j |a_j|$.

\vskip 0.2cm

To establish the above statement, noting that for any $(a_0,...,a_{d-1})\in \mathbb C^d,$ which need not to be normalized, the triangle inequality leads to
\begin{align*}
\Big |\sum_{j\in \mathbb Z_{d}} e^{-i\theta_j}a_j\Big | &\geq \max_k\Big ( |e^{-i\theta_k}a_k| - \sum_{j\in\mathbb Z_d, j\neq k}|e^{-i\theta_j}a_j| \Big ) \\
&= \max_k\Big ( |a_k| - \sum_{j\in\mathbb Z_d, j\neq k}|a_j| \Big ).
\end{align*}
Since
$$
2 \max_j |a_j| - \sum_{j\in \mathbb Z_d} |a_j| = \max_k \Bigl( |a_k| - \sum_{j\in\mathbb Z_d, j\neq k} |a_j| \Bigr),
$$
we have 
$$ S_{\min}(|\psi\rangle) \geq \frac{1}{d} \Big(\max\Big \{0,  2 \max_j |a_j| - \sum_{j\in \mathbb Z_d} |a_j| \Big\}\Big)^2.$$

What remains now is to show that the above inequality is actually saturated. Let $ j_0 $ denote the index of the largest component, i.e., $|a_{j_0}|\geq |a_j|$ for any $j\neq j_0.$ If there are multiple components with the same largest magnitude, choose one of them to be $j_0$. If $ |a_{j_0}| - \sum_{j \neq j_0} |a_j| > 0 ,$ setting 
$$\boldsymbol\theta=e^{i\theta}(\gamma _{0},\cdots, \gamma _{j_0-1},\gamma _{j_0}+\pi, \gamma _{j_0+1},\cdots, \gamma _{d-1} )$$ 
leads to 
\begin{eqnarray*}
\Big |\sum_{j\in \mathbb Z_{d}} e^{-i\theta_j}a_j\Big | &=& \Big |\sum_{j\neq j_0} |a_j| -|a_{j_0}|\Big |\\
&=&|a_{j_0}| - \sum_{j\in\mathbb Z_d,  j\neq j_0} |a_j|,
\end{eqnarray*}
which shows that the inequality is actually an equality.


If $ \max_{k\in\mathbb Z_d} \bigl( |a_j| - \sum_{j \neq k} |a_j| \bigr) \leq 0 $, we need to prove that $S_{\min}(|\psi\rangle)=0$.
We will do this by induction on the dimension. For $d = 2$,
\begin{align*}
S_{\min}(|\psi\rangle) &= \min_{\theta_0, \theta_1} \frac{1}{2} \bigl| e^{-i\theta_0} a_0+ e^{-i\theta_1} a_1 \bigr|^2 \\
&= \frac{1}{2} \bigl( |a_0| - |a _1| \bigr)^2.
\end{align*}
Therefore for the qubit case, Eq. \eqref{Smin} holds. 
Now, assume that Eq. \eqref{Smin} holds for $ d-1 $, in other words, there exists  $\boldsymbol\theta\in[0,2\pi)^{d-1}$ such that 
\begin{equation}\label{assume}
\min_{\boldsymbol\theta} \Bigl| \sum_{j\in \mathbb Z_{d-1}} e^{-i\theta_j} a_j \Bigr| = \max \Big\{ 0, 2 \max_{j \in \mathbb Z_{d-1}} |a_j| - \sum_{j\in \mathbb Z_{d-1}} |a_j|  \Big\}.
\end{equation}
For the $d$-dimensional case, consider the $d-1$ components $a_j$ for $j \in A:= Z_{d} \setminus \{j_0\},$ then
$$
\max \Big \{ 0, 2\max_{j\in A}  |a_j| - \sum_{j\in A } |a_j|  \Big \} \leq \Big |\sum_{j\in A} e^{-i\theta_j} a_j\Big | \leq \sum_{j\in A } |a_j|,
$$
with both bounds being achievable, and when $\boldsymbol\theta $ varies, $|\sum_{j\in A} e^{-i\theta_j} a_j|$ can take any value between the lower and upper bounds. The upper bound is attained by setting $\theta_j=\gamma _j+\theta$ with arbitrary $\theta\in[0,2\pi)$ for any $j\in A.$ The lower bound is achieved by Eq. (\ref{assume}) in the induction. Since 
$$
\max\Big \{ 0, 2\max_{j\in A}  |a_j| - \sum_{j\in A } |a_j|  \Big \} \leq |a_{j_0}| \leq \sum_{j\in A } |a_j|,
$$
 there exists $\boldsymbol\theta$ such that 
$
|\sum_{j \in A} e^{-i\theta_j} a_j| = |a_{j_0}|.
$
Thus in this case we have $$ S_{\min}(|\psi\rangle) =\frac{1}{d} \min_{\boldsymbol\theta} \Big| \gamma _{j_0}- \sum_{j \in A} e^{-i\theta_j} a_j \Big|^2 = 0, $$ which completes the proof.

\vskip 0.1cm

Proposition  7 reveals that $S_{\min}(|\psi\rangle)$ quantifies the polarization extent of the state $|\psi\rangle$ with respect to the computational basis. It is nonzero only when a single component dominates, i.e., $
|a_{j_0}| > \sum_{j \neq j_0} |a_j|.
$
In this case, the state is significantly aligned with a specific basis state $|j_0\rangle$. Conversely, if no such dominant component exists, the unavoidable quantum superposition captured by $S_{\min}$ vanishes. Thus, $S_{\min}$ serves as an indicator of how much a state can be made classical (i.e., close to a basis state) by an optimal choice of the phase parameters.

\subsection{Maximal superposition}

The maximal value of phase-sensitive superposition 
\begin{equation}
S_{\max}(\rho):=\max_{\boldsymbol\theta} \langle \boldsymbol{\theta}|\rho|\boldsymbol{\theta}\rangle
\end{equation}
is a desirable quantifier of superposition as a quantum resource, as first studied in Ref. \cite{Chan2017}. Clearly,
$$S_{\max}(\rho)=\max_{\boldsymbol\theta }  \langle \boldsymbol{0}| U_{\boldsymbol \theta} \rho U_{\boldsymbol \theta}^\dag|{\boldsymbol 0} \rangle.$$

Let $\rho = |\psi \rangle \langle \psi|$ with $|\psi\rangle $ defined by Eq. (\ref{pure}),
then
\begin{equation}\label{Smax}
 S_{\max}(|\psi\rangle)=\frac1d \sum_{j,k\in\mathbb Z_d} |\bar a_j a_k| =\frac1d \sum_{ j\neq k} |\bar a_j a_k|+\frac1d,
\end{equation}
which is essentially the coherence of the state $|\psi\rangle $ \cite{Coh3}. 

\vskip 0.2cm

{\bf Proposition 8}. The maximal superposition $S_{\max}(\cdot)$ has the following properties.

(1) (Boundedness). $1/d\leq S_{\rm max}(\rho)\leq 1.$ $S_{\rm max}(\rho)= 1$ if and only if $\rho$ is a maximally superposed state, while 
$S_{\rm max}(\rho)=1/d$ if and only if $\rho$ is diagonal in the base $\{|j\rangle:j\in\mathbb Z_d\}.$ In particular, for the maximal mixed state ${\bf 1}/d,$ $S_{\rm max}({\bf 1}/d )=1/d.$


(2) (Convexity).  $S_{\max}(\cdot)$ is convex in the sense that
$$S_{\rm max}\Big (\sum_i p_i\rho_i\Big )\leq \sum_i p_iS_{\rm max}(\rho_i)$$  
 for any probability $\{p_i\}$ and quantum states $\rho_i.$

(3) (Tensor product bound). For any quantum states $\rho_a$ and $\rho_b$ on two systems $a$ and $b$,  
$$S_{\rm max}(\rho_a \otimes \rho_b) \geq S_{\rm max}(\rho_a)S_{\rm max}(\rho_b).$$

(4) (Invariance). The maximal superposition is invariant under the diagonal unitary transformation in the sense that
$$S_{\rm max}(U\rho U^\dag)=S_{\rm max}(\rho )$$
for any unitary transform $U$ of the form defined in Eq. (\ref{DU}).

\vskip 0.2cm

Some of the above statements already appeared in Ref. \cite{Chan2017}, for completeness, we review the simple proof. The upper bound in item (1) follows directly form the definition, while the lower bound is a direct consequence of Proposition 3.

Item (2) follows from 
\begin{align*}
&S_{\rm max}\Big (\sum_i p_i\rho_i \Big )=\max_{\boldsymbol\theta} \sum_i p_i\langle \boldsymbol{\theta}|\rho_i|\boldsymbol{\theta}\rangle\\
&\leq \sum_i p_i \max_{\boldsymbol\theta} \langle \boldsymbol{\theta}|\rho_i|\boldsymbol{\theta}\rangle  =\sum_i p_i S_{\rm max}(\rho_i).
\end{align*}

Item (3) follows from
\begin{align*}
\max_{\boldsymbol\theta}  \langle \boldsymbol{\theta}|\rho_a \otimes \rho_b|\boldsymbol{\theta}\rangle &\geq \max_{{\boldsymbol\theta} _a, {\boldsymbol\theta} _b} \langle \boldsymbol{\theta}_a \otimes \boldsymbol{\theta}_b|\rho_a \otimes \rho_b|\boldsymbol{\theta}_a \otimes \boldsymbol{\theta}_b\rangle \\
&= \max_{{\boldsymbol\theta} _a} \langle \boldsymbol{\theta}_a |\rho_a |\boldsymbol{\theta}_a \rangle \max_{{\boldsymbol\theta} _b} \langle \boldsymbol{\theta}_b |\rho_b |\boldsymbol{\theta}_b \rangle.
\end{align*}

Item (4) follows from
 $U^\dag \boldsymbol\Theta  U=\boldsymbol\Theta$ for any unitary transform $U$ defined by Eq. (\ref{DU}).

A tighter upper  bound of the maximal superposition is as follows.
For any quantum state $\rho$ whose largest eigenvalue is $\lambda_M,$ it holds that 
$$S_{\text{max}}(\rho) \leq \lambda_M.$$
The upper bound is achieved if and only if the eigenstate corresponding to $\lambda_M$ is a maximally superposed state.
To prove this, let the spectral decomposition of $\rho$ be
$$
\rho = \sum_{j\in \mathbb Z_d} \lambda_j |\psi_j\rangle \langle \psi_j|,
$$
where $\lambda_0 \geq \lambda_1 \geq \cdots \geq \lambda_{d-1}$ are the eigenvalues of $\rho$, then 
$$
\max_{\boldsymbol\theta} \langle \boldsymbol{\theta} | \rho | \boldsymbol{\theta} \rangle = \max_{\boldsymbol\theta} \sum_{j\in \mathbb Z_d} \lambda_j |\langle \boldsymbol{\theta} | \psi_j \rangle|^2 \leq \lambda_0.
$$
The equality holds if and only if $|\psi_0\rangle\in\boldsymbol\Theta.$ In that case, we can choose $|\boldsymbol{\theta}\rangle = |\psi_0\rangle$.

Since in general, the extreme superposition is not easy to evaluate, it is desirable to obtain various bounds of these quantities. 
Let
\begin{align*}
S_{\rm max}'(\rho)=\max_{k\in\mathbb Z_d} S_k(\rho),\qquad 
S_{\rm min}'(\rho)&=\min_{k\in\mathbb Z_d} S_k(\rho),
\end{align*}
where 
$$S_k(\rho)=\langle \boldsymbol{0}|Z^{\dag k} \rho Z^{ k}|\boldsymbol{0}\rangle,\qquad k\in\mathbb Z_d,$$ and $Z={\rm diag} (1,\omega,\cdots, \omega ^{d-1})$ with $\omega =e^{2\pi i/d}$ is the diagonal $Z$-gate.
Noting that $\{Z^k|\boldsymbol{0}\rangle: k\in \mathbb{Z}_d\}$ constitutes an orthonormal basis of $\mathbb{C}^d$, we have 
$\sum_{k\in\mathbb Z_d} S_k(\rho)=1.$
Obviously, $$S_{\rm min}(\rho)\leq S_{\rm min}'(\rho)\leq S_{\rm max}'(\rho)\leq S_{\rm max}(\rho).$$
Therefore, the maximal and minimal superposition can be approximated by these easily measurable values.

\section{Conclusion} 

In this work, we have quantified the phase-sensitive superposition directly based on the overlap between the concerned quantum state and the reference state, which is an arbitrary maximally superposed state. We have highlighted the role played by various phases. We have found that the average of phase-sensitive superposition is a constant number independent of the quantum state, which implies that there is a conservation relation for phase-sensitive superposition. The physical meaning of this conservation relation lies in the complementary relation between superposition for different phases.
It is quite natural to generalize the overlap to any form of similarity measures between the concerned quantum state and the reference state, such as affinity \cite{Luo2004}.

We have studied various aspects of phase-sensitive superposition, including its relations to quantum design, complete decoherence channel and quantum coherence. In particular, setting $|\boldsymbol \theta\rangle=|\boldsymbol 0\rangle,$ the quantifier of superposition $S_{\boldsymbol 0} $ is closely related to the recently introduced notion  ``quantum state texture" \cite{Pari2024}, which characterize the irregularity of a quantum  operator in a selected basis. Consequently, the phase-sensitive superposition generalizes the quantum state texture to arbitrary phases and moreover exploits the freedom of phase to characterize quantum states from the perspective of superposition.

We have further studied  the extreme values (minimal and maximal) of phase-sensitive superposition, which have desirable properties, though hard to be analytically calculated in general due to the complex optimization, we have derived analytical expressions of extreme superposition for and pure state, as well as  some simple bounds for mixed states. 

 Applying these quantifiers of superposition in Grover search algorithm (see the Appendix), we have established a trade-off relation between the maximal superposition and success probability, which  provides an operational illustration
of de-superposition as success probability. 

It is desirable to further investigate implications and role of phase-sensitive superposition in quantum information processing tasks, and study its interplay with coherence, entanglement, and other quantum resources.

\vskip 0.4cm
\noindent {\bf Acknowledgements}.
This work was supported by the National Natural Science Foundation of China (Grant Nos. 12426671, 12401609,  and 12341103), the Youth Promotion Association of CAS (Grant No. 2023004), and the Beijing Natural Science Foundation (Grant No. Z250004).

\vskip0.3cm

\noindent
\begin{center}
    \textbf{APPENDIX}
\end{center}

In this Appendix, we first illustrate our quantifiers of superposition by evaluating the phase-sensitive superposition of several paradigmatic states (Appendix A). We then discuss an application of these quantifiers to the Grover search algorithm, characterize the dynamics of superposition throughout the computational process, and  reveal its relation to the success probability of the search algorithm  (Appendix B).

\vspace{0.2cm}

\begin{center}
    \textbf{A. Illustrative examples}
\end{center}

{\bf Example 1}.  For a qubit system $\mathbb{C}^2$  with canonical basis $\{|0\rangle, |1\rangle \},$ any  state $\rho=(\rho_{jk})$ has the Bloch representation 
$\rho=\big  ({\bf 1}+\sum_{j=1}^3 r_j\sigma_j\big )/2$
where $r_j\in[-1,1],$ $\sum_{j=1}^3 r_j^2\leq1,$ and $\sigma_j$ are  the Pauli matrices. Direct calculation leads to 
\begin{equation*}
S_{\boldsymbol{\theta}}(\rho)=\frac12\big(1-r_1 \cos(\theta_1-\theta_0)+r_2 \sin(\theta_1-\theta_0)\big),
\end{equation*}
from which we obtain
\begin{align*}
S_{\min} (\rho)&=\frac12\Big(1-\sqrt{r_1^2+r_2^2}\Big) =\frac12-|\rho_{12}|,\\
S_{\max} (\rho)&=\frac12\Big(1+\sqrt{r_1^2+r_2^2}\Big) =\frac12+|\rho_{12}|.
\end{align*}
It is interesting to note that 
$$S_{\min} (\rho)+S_{\max} (\rho)=1.$$
However, this  does not hold for $d>2$ as indicated by Eqs.  (\ref{Smin}) and (\ref{Smax}). 

\vskip 0.2cm

{\bf Example 2}.  From a qubit system $\mathbb{C}^2$  with canonical basis $\{|0\rangle, |1\rangle \},$ we can construct the tensor product system $\mathbb{C}^2\otimes \mathbb{C}^2=\mathbb{C}^4$ with basis $\{|j\rangle =|j_0j_1\rangle : j\in \mathbb{Z}_4\}$ ($j=j_0+j_1\cdot 2$ in binary expansion). For the Bell states 
\begin{equation*}
|\Phi_\pm\rangle=\frac1{\sqrt2} (|00\rangle\pm|11\rangle),\quad |\Psi_\pm\rangle=\frac1{\sqrt2} (|01\rangle\pm|10\rangle),
\end{equation*}
then various quantities concerning superposition can be straightforwardly evaluated as 
\begin{align*}
  S_{\boldsymbol{\theta}}(|\Phi_\pm \rangle) &= \frac18 \, \bigl| e^{-i\theta_0} \pm e^{-i\theta_3} \bigr|^2, \\
S_{\boldsymbol{\theta}}(|\Psi_\pm \rangle) &= \frac18 \, \bigl| e^{-i\theta_1} \pm  e^{-i\theta_2} \bigr|^2, \\
S_{\boldsymbol 0}(|\Phi_+\rangle)&=S_{\boldsymbol 0}(|\Psi_+\rangle)=\frac12,\\
S_{\boldsymbol 0}(|\Phi_-\rangle)&=S_{\boldsymbol 0}(|\Psi_-\rangle)=0,\\
S_{\min}(|\Phi_\pm\rangle)&=S_{\min}(|\Psi_\pm\rangle)=0,\\
S_{\rm max}(|\Phi_\pm\rangle)&=S_{\rm max}(|\Psi_\pm\rangle)=\frac12.
\end{align*}

For the Werner states $\rho_p=p|\Psi_-\rangle\langle\Psi_-|+(1-p){\bf 1}/4$ on $\mathbb{C}^4,$ we have
\begin{align*}
&S_{\boldsymbol\theta}(\rho_p)=\frac{1-p}{4} + \frac{p}{8} \bigl| e^{-i\theta_1} - e^{-i\theta_2} \bigr|^2, \ \ S_{\boldsymbol 0}(\rho_p)=\frac{1-p}{4},\\
&S_{\min}(\rho_p)= \frac{1-p}{4},\qquad S_{\rm max}(\rho_p)=\frac{1+p}4,
\end{align*}
and for the isotropic state $\tau_p=p|\Psi_+\rangle\langle\Psi_+|+(1-p){\bf 1}/4$ on $\mathbb{C}^4,$ we have
\begin{align*}
&S_{\boldsymbol\theta}(\tau_p)=\frac{1-p}{4} + \frac{p}{8} \bigl| e^{-i\theta_1} + e^{-i\theta_2} \bigr|^2,\ \ S_{\boldsymbol 0}(\tau_p)= \frac{1+p}{4},\\
&S_{\rm min}(\tau_p)=\frac{1-p}{4},\qquad S_{\rm max}(\tau_p)=\frac{1+p}{4}.
\end{align*}

{\bf Example 3}. From a qubit system $\mathbb{C}^2$  with canonical basis $\{|0\rangle, |1\rangle \},$ we can construct the tensor product system $\mathbb{C}^2\otimes \mathbb{C}^2\otimes \mathbb{C}^2=\mathbb{C}^8$ with basis $\{|j\rangle =|j_0j_1j_2\rangle : j\in \mathbb{Z}_8\}$ ($j=j_0+j_1\cdot 2+j_2\cdot 2^2$ in the binary expansion). 
For the W states $|{\rm W}\rangle=\frac1{\sqrt3} (|001\rangle+|010\rangle+|100\rangle)\in \mathbb{C}^8,$  it holds that
\begin{align*}
&S_{\boldsymbol \theta}(|{\rm W}\rangle) =\frac{1}{24}|e^{i\theta_1}+e^{i\theta_2}+e^{i\theta_4}|^2,\ \ S_{\boldsymbol 0}(|{\rm W}\rangle)=\frac38,\\
&S_{\rm min}(|{\rm W}\rangle)=0,\qquad S_{\rm max}(|{\rm W}\rangle)=\frac38.
\end{align*}

For $W_p=(1-p)|{\rm W}\rangle\langle {\rm W}|+p\frac{\bf 1}{8},$ we have
\begin{align*}
&S_{\boldsymbol \theta}(W_p)=\frac{p}{8}+\frac{1-p}{24}|e^{i\theta_1}+e^{i\theta_2}+e^{i\theta_4}|^2, \\
& S_{\boldsymbol 0}(W_p)=\frac{3-2p}{8},\ \  S_{\rm min}(W_p)=\frac{p}{8},\ \  S_{\rm max}(W_p)=\frac{3-2p}{8}.
\end{align*}

For the GHZ states $|{\rm GHZ}\rangle=\frac1{\sqrt2} (|000\rangle+|111\rangle),$ we have
\begin{align*}
&S_{\boldsymbol \theta}(|{\rm GHZ}\rangle)=\frac{1}{16}|e^{i\theta_0}+e^{i\theta_7}|^2,\quad S_{\boldsymbol 0}(|{\rm GHZ}\rangle)=\frac14,\\
&S_{\rm min}(|{\rm GHZ}\rangle)=0,\qquad S_{\rm max}(|{\rm GHZ}\rangle)=\frac14.
\end{align*}

For $G_p=(1-p)|{\rm GHZ}\rangle\langle {\rm GHZ}|+p\frac{\bf 1}{8},$ we have
\begin{align*}
&S_{\boldsymbol \theta}(G_p)=\frac{p}{8}-\frac{1-p}{16}|e^{i\theta_0}+e^{i\theta_7}|^2,\quad S_{\boldsymbol 0}(G_p)=\frac{2-p}{8},\\
&S_{\rm min}(G_p)=\frac{p}{8},\qquad S_{\rm max}(G_p)=\frac{2-p}{8}.
\end{align*}

For the Bell-diagonal states $\rho = (\mathbf{1} + \sum_{j} c_j \sigma_j \otimes \sigma_j)/4,$ we have
\begin{align*}
&S_{\boldsymbol{\theta}}(\rho) = \frac14 + \Big ( \frac{c_1 + c_2}{8} x + \frac{c_1 - c_2}{8} y \Big ), \ \ S_{\boldsymbol0}(\rho) = \frac{1 + c_1}{4}, \\
&S_{\min}(\rho) = \frac 14 (1 - \max(|c_1|, |c_2|)), \\
&S_{\max}(\rho) = \frac 14 (3 - \max(|c_1|, |c_2|)),
\end{align*}
where $x = \cos(\theta_3 - \theta_0), \ y = \cos(\theta_2 - \theta_1).$

\vskip 0.2cm

{\bf Example 4}. In a $(2j+1)$-dimensional spin-$j$ system $\mathbb{C}^{2j+1},$  the spin-$j$ coherent states \cite{Per}
\begin{eqnarray*}
|\zeta\rangle= \sum_{m=-j}^j\sqrt{2j \choose j-m}\frac{\zeta^{j-m}}{(1+|\zeta|^2)^j}|j,m\rangle,\qquad \zeta\in\mathbb C,
\end{eqnarray*}
expanded in the Dicke states  $|j,m\rangle,$ which are  defined as the simultaneous eigenstates of $J_z$ and the total spin observable
$\boldsymbol{J}^2=J_x^2+J_y^2+J_z^2$ as
\begin{equation*}
J_z|j, m\rangle=m|j,m\rangle, \qquad \boldsymbol{J}^2|j,m\rangle =j(j+1)|j, m\rangle. 
\end{equation*}
The Dicke states constitute an orthonormal basis   $\{|j,m\rangle : m=-j, -j+1, \dots, j-1,j\}$ of $\mathbb{C}^{2j+1}.$ In this case,  the maximally superposed states are 
$$
|\boldsymbol{\theta}\rangle = \frac{1}{\sqrt{2j+1}} \sum_{m=-j}^{j} e^{i\theta_m} |j, m\rangle, \quad \  \boldsymbol{\theta}=(\theta _{-m}, \dots, \theta_m)
$$
and we have
$$
\begin{aligned}
   &  S_{\boldsymbol{\theta}}(|\zeta\rangle)=  \langle\boldsymbol{\theta}|\zeta\rangle\langle\zeta|\boldsymbol{\theta}\rangle \\ 
    &=\frac{1}{(2j+1) (1 + |\zeta|^2)^{2j}} \Bigg | \sum_{m=-j}^{j} \sqrt{\binom{2j}{j+m}}  \zeta^{j-m} e^{-i\theta_m} \Bigg |^2.
\end{aligned}
$$
In particular,  
$$
\begin{aligned}
   S_{\mathbf{0}}(|\zeta\rangle) =  \frac{1}{(2j+1) (1 + |\zeta|^2)^{2j}} \Bigg | \sum_{m=-j}^{j} \sqrt{\binom{2j}{j+m}}  \zeta^{j-m} \Bigg |^2.
\end{aligned}
$$

To obtain $S_{\max}(|\zeta\rangle)$, we can adjust $\theta_m$ so that all terms in the sum are in phase and contribute positively, which yields
$$
\begin{aligned}
    &S_{\max}(|\zeta\rangle)= \max_{\boldsymbol{\theta}} \langle \boldsymbol{\theta}|\zeta\rangle\langle\zeta|\boldsymbol{\theta}\rangle \\ 
    &= \frac{1}{(2j+1) (1 + |\zeta|^2)^{2j}} \Bigg ( \sum_{m=-j}^{j} \sqrt{\binom{2j}{j+m}}  |\zeta|^{j-m} \Bigg )^2.
\end{aligned}
$$
For $S_{\min}(|\zeta\rangle)$, it follows from Eq. \eqref{Smin} that 
$$
\begin{aligned}
&S_{\min}(|\zeta\rangle)=\min_{\boldsymbol{\theta}}\langle \boldsymbol{\theta}|\zeta\rangle\langle\zeta|\boldsymbol{\theta}\rangle = \\
& \begin{cases}
\dfrac{\Big(2 \max\limits_{m} a_m - \sum\limits_{m=-j}^{j} a_m\Big)^2}{(2j+1)(1+|\zeta|^2)^{2j}}, & \text{if } 2 \max\limits_{m} a_m \ge \sum\limits_{m=-j}^{j} a_m, \\
0, & \text{otherwise},
\end{cases}
\end{aligned}
$$
where
$$
a_m = \sqrt{\binom{2j}{j+m}} |\zeta|^{j-m}.
$$


\vskip 0.1cm

\begin{center}
    \textbf{B. Superposition in Grover search algorithm}
\end{center}

Here we investigate the dynamics of superposition in Grover search algorithm in order to reveal its quantum feature. First, we recall the  Grover search algorithm \cite{Gro1,Grov19971,Gro2,Sun2024}. 
In an unstructured database $\mathcal N$ with $N$ items, represented by an register with the computational basis $\{|j\rangle: j\in\mathbb Z_N\}$ of $\mathbb{C}^N$,  let $\mathcal M\subset \mathcal N$ be the set of $M$ marked items, and 
    $\mathcal M^\perp= \mathcal N\setminus \mathcal M$ be the set of unmarked items. If the initial state of the database is 
$$|\varphi_0\rangle=\frac1{\sqrt N}\sum_{j\in\mathbb Z_N} |j\rangle,$$
which is equal to $|\boldsymbol 0\rangle$ defined by Eq. (\ref{zero}), then it can be rewritten as
$$|\varphi_0\rangle=\sin\frac\alpha2|\mathcal M\rangle+\cos\frac\alpha2|\mathcal M^\perp\rangle $$
with $\alpha =2{\rm arc sin} \sqrt{\frac MN}$ and 
$$|\mathcal M\rangle =\frac1{\sqrt M}\sum_{j\in\mathcal M} |j\rangle, \quad |\mathcal M^\perp\rangle =\frac1{\sqrt{N-M}}\sum_{j\in\mathcal M^\perp} |j\rangle.$$

After $t$ iterations of performing the Grover operation $DO$ on the initial state $|\varphi_0\rangle $, where 
$$D=2|\varphi_0\rangle\langle \varphi_0|-{\bf 1}, \quad O={\bf 1}-2|\mathcal M\rangle\langle \mathcal M |$$
are the inversion operation and  the oracle unitary operation, respectively, the $t$-th time state is 
$$|\varphi_t\rangle=(DO)^t|\varphi_0\rangle= \sin\Big(\frac\alpha2+t\alpha\Big)|\mathcal M\rangle+\cos\Big(\frac\alpha2+t\alpha\Big)|\mathcal M^\perp\rangle.$$

The success probability after $t$ iterations for finding a marked state by making a projective measurement $\{\Pi _{\cal M}, \Pi _{{\cal M}^\perp}\}$ (here $\Pi _{\cal M}=\sum _{j\in {\cal M}}|j\rangle \langle j|$) on $|\varphi_t\rangle$ is 
\begin{equation}\label{pt}
P(|\varphi _t\rangle )=\sum_{j\in\mathcal M} {\rm tr} \big (|\varphi_t\rangle \langle\varphi_t| j\rangle\langle j|\big )=\sin^2\Big(\frac\alpha2+t\alpha\Big).
\end{equation}
For the first time when $P_t$ approaches 1, $(1/2+t)\alpha\in[0,\pi/2]$ and the optimal number of iterations for maximizing the success probability is 
$$t_{\rm opt}= \Big \lfloor \frac\pi 4 \sqrt{\frac NM}\Big \rfloor,$$
the largest integer upper bounded  by $\frac\pi 4 \sqrt{\frac NM}.$

By direct calculations, we obtain 
\begin{align*}
S_{\boldsymbol \theta}(|\varphi_t\rangle)=&\Big|\sin\frac\alpha2\sin\Big(\frac\alpha2+t\alpha\Big)\Big|^2S_{\boldsymbol \theta_{\mathcal M}}(|\mathcal M\rangle)\\
&+\Big|\cos\frac\alpha2\cos\Big(\frac\alpha2+t\alpha\Big)\Big|^2S_{\boldsymbol \theta_{\mathcal M^\perp}}(|\mathcal M^\perp\rangle) \\
&+\frac12\sin\alpha\sin\big((1+2t)\alpha\big){\rm Re} X,
\end{align*}
where $X=\langle \mathcal M|\boldsymbol\theta_{\mathcal M}\rangle \langle\boldsymbol\theta_{\mathcal M^\perp}|\mathcal M^\perp\rangle$ and  
\begin{align*}
|\boldsymbol \theta\rangle&=\sin\frac\alpha2|\boldsymbol\theta_{\mathcal M}\rangle+\cos\frac\alpha2|\boldsymbol\theta_{\mathcal M^\perp}\rangle, \\
|\boldsymbol\theta_{\mathcal M}\rangle &= \frac1{\sqrt M}\sum_{j\in \mathcal M} e^{i\theta_j} |j\rangle,\\
|\boldsymbol\theta_{\mathcal M^\perp}\rangle &= \frac1{\sqrt {N-M}}\sum_{j\in \mathcal M^\perp} e^{i\theta_j} |j\rangle.
\end{align*}

Obviously, the maximal value of phase-sensitive superposition is
\begin{align}
\nonumber & S_{\max} (|\varphi_t\rangle)=\Big|\sin\frac\alpha2\sin\Big(\frac\alpha2+t\alpha\Big)\Big|^2\\
\nonumber & +\Big|\cos\frac\alpha2\cos\Big(\frac\alpha2+t\alpha\Big)\Big|^2 +|\frac12\sin\alpha\sin\big((1+2t)\alpha\big)|\\
&=\max \big \{\cos^2 (t\alpha), \cos^2 (t\alpha+\alpha)\big\},\label{gsmax}
\end{align}
which is attained when 
\begin{align*}
&S_{\boldsymbol \theta_{\mathcal M}}(|\mathcal M\rangle)=S_{\boldsymbol \theta_{\mathcal M^\perp}}(|\mathcal M^\perp\rangle)=1,\\
&\sin\alpha\sin\big((1+2t)\alpha\big){\rm Re} X=\big |\sin\alpha\sin\big((1+2t)\alpha\big) \big |.
\end{align*}
The above conditions are satisfied if we take  $|\boldsymbol\theta_{\mathcal M}\rangle=|\mathcal M\rangle$ and $|\boldsymbol\theta_{\mathcal M^\perp}\rangle=-|\mathcal M^\perp\rangle$ when $\sin\alpha\sin\big((1+2t)\alpha\big)<0,$ and
take $|\boldsymbol\theta_{\mathcal M}\rangle=|\mathcal M\rangle$ and $|\boldsymbol\theta_{\mathcal M^\perp}\rangle=|\mathcal M^\perp\rangle$ when $\sin\alpha\sin\big((1+2t)\alpha\big)\geq0.$ In the later case, $S_{\max} (|\varphi_t\rangle)=S_{\boldsymbol 0} (|\varphi_t\rangle).$

When $M \geq 2$ and $N - M \geq 2$, the minimal value of the phase-sensitive superposition over $\boldsymbol\Theta$ is
$$
S_{\min} (|\varphi_t\rangle) = 0,
$$
which holds when $\langle \mathcal{M}|\boldsymbol\theta_{\mathcal{M}}\rangle = \langle \boldsymbol\theta_{\mathcal{M}^\perp}|\mathcal{M}^\perp\rangle = 0$.

When $ M = 1 $ or $ N - M = 1 $, the global minimum value $ S_{\min} (|\varphi_t\rangle)$ is not necessarily zero. To illustrate this, consider the case where $ M = 1 $ and $ N \gg M $, which can occur in practical situations. According to Proposition 7, $S_{\min}(|\varphi_t\rangle) $ equals to
$$\frac{1}{N} \Big( \max \Big\{0,  \Big |\sin \Big (\frac\alpha2+t\alpha\Big )\Big | - \sqrt{N-1}\Big |\cos\Big (\frac\alpha2+t\alpha\Big )\Big | \Big\}\Big )^2.$$
The maximum value of $ S_{\min}(|\varphi_t\rangle) $ occurs when $ t $ corresponds to the highest probability of success.

Comparing Eqs. (\ref{pt}) and (\ref{gsmax}), we obtain  a complementary relation between
superposition and success probability.

\vskip 0.2cm

{\bf  Proposition  9.} For the Grover search algorithm, the maximum superposition of the evolved state $|\varphi_t\rangle$ and the success probability $P(|\varphi _t\rangle )$ satisfy 
$$S_{\max}(|\varphi_t\rangle )=\bigg(\sqrt{\big (1-P(|\varphi _t\rangle )\big )\Big (1-\frac MN\Big)}+\sqrt{P(|\varphi _t\rangle )\frac MN} \bigg)^2,$$
where $(1/2+t)\alpha\in[0,\pi/2]$ with $\alpha =2{\rm arc sin} \sqrt{\frac MN}.$
In particular, for $N\gg M,$ we have
$$S_{\max}(|\varphi_t\rangle )+P(|\varphi _t\rangle )\simeq 1.
$$

\vskip 0.2cm
 
We depict the dynamics of the maximal superposition and the success probability of the evolved search state $|\varphi _t\rangle $ in Fig. 1, which exhibits clearly the complementary relation between superposition and success probability and implies that higher success probability requires more consumption of superposition.

\begin{figure}
    \centering    \includegraphics[width=0.9\linewidth]{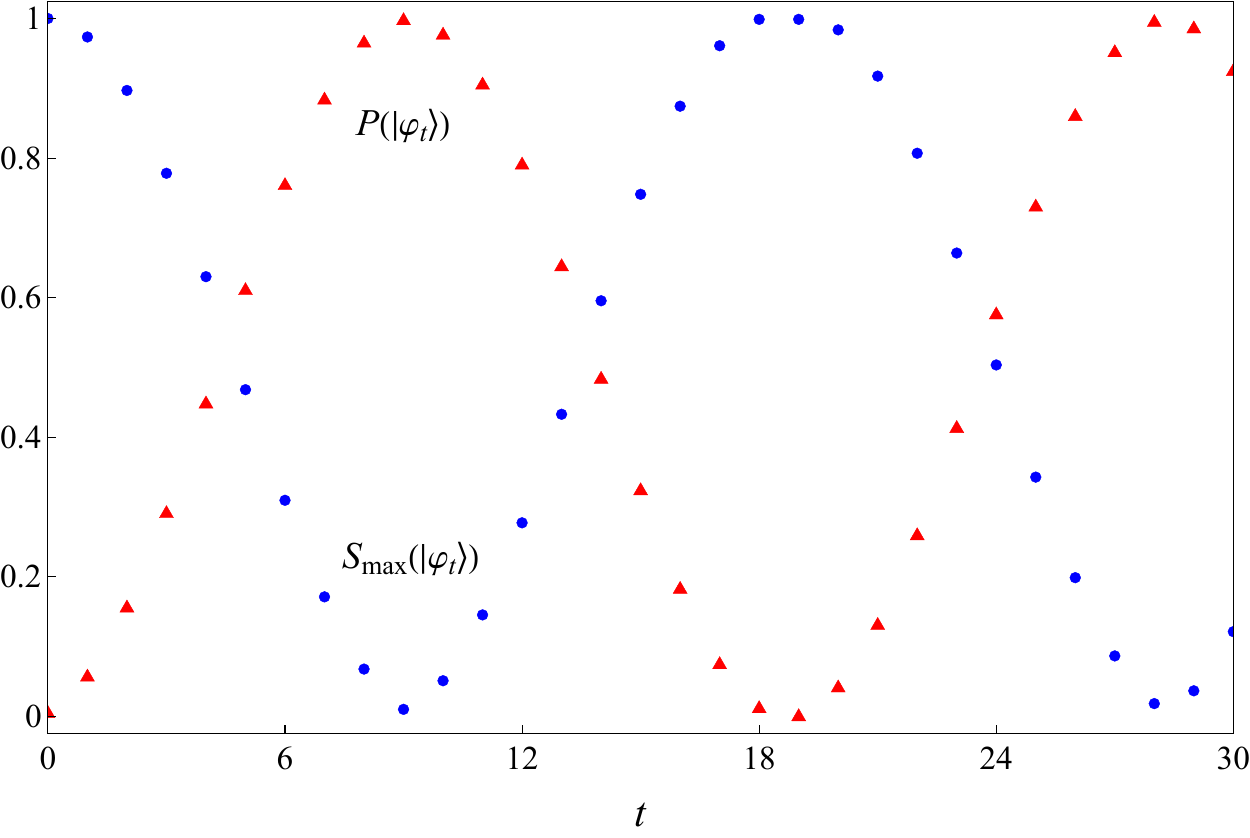}
    \caption{Dynamics of the maximal superposition $S_{\max}(|\varphi_t\rangle )$ and the success probability  $P_{\max}(|\varphi_t\rangle )$ of the evolved state $|\varphi_t\rangle.$ Here  $t$ denotes the number of iterations, $M=1,N=150.$ We see clearly that the superposition is complementary to the success probability, or equivalently, the depletion of superposition is proportional to the success probability: The Grover search algorithm consumes superposition.}
    \label{fig1}
\end{figure}

The above approximate equality relation between superposition and success probability also holds in  Grover search algorithm with more general initial superposition states. To illustrate this, take the initial state
$$|\psi_0\rangle=\sin\frac\beta2|\mathcal M\rangle+\cos\frac\beta2 e^{i\phi}|\mathcal M^\perp\rangle,\quad \beta\in[0,\pi],\phi\in[0,2\pi), $$
then the $t$-th time state is 
\begin{align*}
|\psi_t\rangle= (DO)^t|\psi _0\rangle=  s_t|\mathcal M\rangle+c_t|\mathcal M^\perp\rangle,
\end{align*}
where 
\begin{align*}
s_t&=\sin\frac\beta2\cos (t\alpha )+e^{i\phi}\cos\frac\beta2\sin (t\alpha ),\\
c_t&= \cos\frac\beta2\cos (t\alpha )-e^{i\phi}\sin\frac\beta2\sin (t\alpha ),\\
\alpha &= 2{\rm arc sin} \sqrt{\frac MN}.
\end{align*}

The success probability is
$$P(|\psi _t\rangle )=|s_t|^2=\Big|\sin\frac\beta2\cos (t\alpha )+\cos\frac\beta2\sin (t\alpha ) e^{i\phi}\Big|^2.$$

The phase-sensitive superposition and the maximal superposition are
\begin{align*}
S_{\boldsymbol \theta}(|\psi_t\rangle)&=
\Big|\sin\frac\alpha2 s_t\Big|^2S_{\boldsymbol \theta_{\mathcal M}}(|\mathcal M\rangle)\\
&\ +\Big|\cos\frac\alpha2 c_t\Big|^2S_{\boldsymbol \theta_{\mathcal M^\perp}}(|\mathcal M^\perp\rangle)+\sin\alpha{\rm Re} Y,\\
S_{\max}(|\psi_t\rangle)&=\Big|\sin\frac\alpha2 s_t\Big|^2+\Big|\cos\frac\alpha2 c_t\Big|^2+|s_tc_t\sin\alpha|,
\end{align*}
where $Y=\bar s_tc_t\langle \mathcal M|\boldsymbol\theta_{\mathcal M}\rangle \langle\boldsymbol\theta_{\mathcal M^\perp}|\mathcal M^\perp\rangle.$

For the decoherence in the Grover search algorithm, which is independent of the amplitudes of the initial states, the relation between it and success probability does not hold in the above generalized cases \cite{Sun2024}. In sharp contrast, the superposition of the initial states is related to both the amplitudes and phases of the initial states, and the relation between it and success probability still holds: For the generalized Grover search algorithm mentioned above, the superposition of the evolved state $|\psi_t\rangle$ and the success probability $P(|\varphi _t\rangle )$ satisfy similar relations as in Proposition 9.



\end{document}